\documentclass[aps,prl,twocolumn,superscriptaddress,floatfix,amsmath]{revtex4}

\begin{document}
\title{\bf Partial Survival and Crossing Statistics for a Diffusing Particle 
in a Transverse Shear Flow}
\author{Alan J. Bray}
\affiliation{School of Physics and Astronomy, The University of Manchester, 
Manchester M13 9PL, UK}
\author{Satya N. Majumdar}
\affiliation{Laboratoire de Physique Th\'eorique et
Mod\`eles Statistiques (UMR 8626 du CNRS),
Universit\'e Paris-Sud, B\^at. 100, 91405 Orsay Cedex, France}

\date{\today}

\begin{abstract}
We  consider  a  non-Gaussian  stochastic  process  where  a  particle
diffuses   in  the  $y$-direction,   $dy/dt=\eta(t)$,  subject   to  a
transverse shear  flow in the  $x$-direction, $dx/dt=f(y)$. Absorption
with probability  $p$ occurs  at each crossing  of the line  $x=0$. We
treat  the  class  of  models  defined  by  $f(y)  =  \pm  v_{\pm}(\pm
y)^\alpha$ where the  upper (lower) sign refers to  $y>0$ ($y<0$).  We
show that the particle survives  up to time $t$ with probability $Q(t)
\sim  t^{-\theta(p)}$  and  we   derive  an  explicit  expression  for
$\theta(p)$  in  terms  of  $\alpha$  and the  ratio  $v_+/v_-$.  From
$\theta(p)$  we  deduce  the  mean  and variance  of  the  density  of
crossings of the line $x=0$ for this class of non-Gaussian processes.

\noindent

\medskip\noindent  {PACS  numbers: 02.50.-r, 05.40.-a}
\end{abstract}

\maketitle



There has been  a resurgence of interest in  first-passage problems in
recent years \cite{Guide}, particularly in the context of systems with 
many degrees of freedom \cite{SatyaReview}.  
There are, however, relatively few  exactly solved models
involving {\em non-Gaussian} processes.  This paper is devoted a class
of such models, involving just two degrees of freedom, for which exact
results can  be obtained.  These models are  most simply  described in
terms of a particle moving  in the two-dimensional plane $(x,y)$, with
stochastic motion  in the $y$-direction and  `deterministic' motion in
the $x$-direction (deterministic in the sense that the velocity in the
$x$-direction  depends only  on  the $y$-coordinate,  which is  itself
however a stochastic variable).

The class of models is defined by the following equations:
\begin{eqnarray}
\label{langevin1}
\dot{y} & = & \eta(t)\ , \\ 
\dot{x} & = & f(y) = v_{\pm} {\rm sgn}(y)\,|y|^\alpha\ ,
\label{langevin2}
\end{eqnarray}
where the  upper (lower) sign  refers to $y>0$ ($y<0$),  dots indicate
time derivatives, and $\eta(t)$ is Gaussian white noise with mean zero
and  correlator  $\langle  \eta(t)\eta(t')\rangle  =  2D\delta(t-t')$.
These  models  are  non-Gaussian   except  for  the  case  $\alpha=1$,
$v_+=v_-=v$,  which   reduces  to  the   random  acceleration  process
$\ddot{x} = v \eta(t)$.

Previous work on this class  of models has addressed the first-passage
problem  in which  the line  $x=0$ is  an absorbing  boundary  and the
particle  starts  in the  half-plane  $x>0$.  It  was shown 
\cite{Bray-Gonos1} that  the probability, $Q(t)$, that the  particle 
survives until time $t$ decays as $Q(t) \sim t^{-\theta}$ for large $t$, 
with
\begin{equation}
\theta            =           \frac{1}{4}-\frac{1}{2\pi\beta}\tan^{-1}
\left[\frac{\gamma - 1}{\gamma + 1}
\tan\left(\frac{\pi\beta}{2}\right)\right]
\label{result}
\end{equation}
where 
\begin{equation}
\label{beta}
\beta = \frac{1}{2+\alpha}, \ \ \ \ 
\gamma = \left(\frac{v_+}{v_-}\right)^\beta. 
\end{equation} 
The exponent $\theta$ takes the value $1/4$ for all $\alpha$ for the 
`antisymmetric' case $\gamma =1 $ (for which $f(y)$ is an odd function) 
which has a simple explanation \cite{RK,Bray-Gonos2} in terms the Sparre 
Anderson theorem \cite{SA}. 

In the  present paper we consider the  case where the line  $x=0$ is a
{\em partially} absorbing boundary,  at which the particle is absorbed
with probability  $p$ at each  crossing. We have shown  elsewhere 
\cite{Majumdar-Bray} that for  such models  the  survival probability  
decays  at late times as a power law with an  exponent $\theta$    that   
depends, in    general,   on    $p$:    $Q(t)   \sim t^{-\theta(p)}$. 
Explicit analytical results for $\theta(p)$ have been obtained for  
a number of  simple (and not-so-simple) models \cite{Majumdar-Bray}. We 
also showed  that the function  $\theta(p)$ determines the moments of the
number of crossings of the boundary  $x=0$ for the case where there is
no absorption.  In the present work we  use the methods of  
ref.~\cite{Bray-Gonos1} to determine the function  $\theta(p)$ and the 
method outlined  in ref.~\cite{Majumdar-Bray}
to  investigate  the  crossing  statistics  for  the  problem  without
absorption, obtaining closed form results for the mean and variance of
the number of crossings in  a given (large) time interval. These results
for the mean and variance are,  to our knowledge,  the first such  
results for non-Gaussian processes.


The derivation  of $\theta(p)$ follows closely that  of $\theta(0)$ in
Ref.~\cite{Bray-Gonos1}, so we only give the main steps. The principal 
difference from the  former  work lies  in  the  boundary  conditions 
imposed  by  the partially absorbing boundary.

From Eqs.\  (\ref{langevin1}) and (\ref{langevin2}) we  can write down
the backward Fokker-Planck equation
\begin{equation}
\frac{\partial Q}{\partial t} = D\frac{\partial^2 Q}{\partial y^2} \pm
v_{\pm} (\pm y)^\alpha \frac{\partial Q}{\partial x}\ ,
\label{BFPE}
\end{equation}
where $Q(x,y,t)$  is the probability that the  particle still survives
at  time  $t$  given that  it  started  at  position ($x$,  $y$).  The
partially absorbing boundary at $x=0$ implies the boundary conditions
\begin{eqnarray}
\label{bc1}
Q(0+,-y,t) & = &  p\,\tilde{Q}(0+,y,t),\ \ \ y>0 \\ 
\tilde{Q}(0+,-y,t) & = & p\,Q(0+,y,t),\ \ \ y>0,
\label{bc2}
\end{eqnarray}
where $\tilde{Q}(x,y,t)$  is the survival  probability for a  model in
which  $v_+$ and  $v_-$ are  interchanged. It  is clear  that  $Q$ and
$\tilde{Q}$ are described by the  same  value  of  $\theta$.  In  fact
$\tilde{Q}(x,y,t) = Q(-x,-y,t)$,  since after interchanging $v_+$ and
$v_-$ the  system is  restored to its  original configuration  after a
rotation by $\pi$ about an axis perpendicular to the $xy$ plane.
The  initial condition  is $Q(x,y,0) = 1 = \tilde{Q}(x,y,0)$.

Solving the full initial value  problem is difficult so we will follow
the approach used in ref.~\cite{Bray-Gonos1} by specializing to the 
late-time scaling regime where $Q(x,y,t) \sim t^{-\theta(p)}$.  This 
approach exploits a generalization   of   the   method   introduced   by   
Burkhardt \cite{Burkhardt}  for
$\alpha=1=\gamma$ (the  random acceleration problem).
The idea  is to  extract explicitly the  time-dependence $t^{-\theta}$
expected at large $t$. This gives, asymptotically,
\begin{equation}
Q(x,y,t)  \sim \left(\frac{x^{2\beta}}{t}\right)^\theta F_\pm\left(\pm
\frac{v_{\pm}(\pm y)^{1/\beta}}{Dx}\right)\ ,
\label{scaling}
\end{equation}
and  a   similar  expression  for   $\tilde{Q}(x,y,t)$,  with  scaling
functions  $\tilde{F}_\pm$.  In  Eq.\ (\ref{scaling}),  the  functions
$F_\pm(z)$ are  the scaling function  for $y>0$ ($+$) and  $y<0$ ($-$)
respectively.  The  (dimensional) prefactors (for $y>0$  and $y<0$) in
Eq.\  (\ref{scaling}) have  been  omitted since  Eq.\ (\ref{BFPE})  is
linear. The functions $F_+(z)$ and  $F_-(z)$ are defined such that the
prefactor is the same for $y>0$ and $y<0$.

Inserting  the form  (\ref{scaling}) into  the  backward Fokker-Planck
equation  (\ref{BFPE}), we  see  immediately that  the term  $\partial
Q/\partial t$  leads to  a term of order  $t^{-(\theta +1)}$, which is
subdominant for large $t$ and  can therefore be dropped. The remaining
terms give
\begin{equation}
z F_\pm''(z) +  (1- \beta  - \beta^2 z) F_\pm'(z) +  2\beta^3 \theta
F_\pm(z)=0\ .
\end{equation}
Expressed in terms of the variable $u=\beta^2z$, this equation becomes
Kummer's   equation.    Independent   solutions  are   the   confluent
hypergeometric   functions   $M(-2\beta\theta,1-\beta,\beta^2z)$   and
$U(-2\beta\theta,1-\beta,\beta^2z)$     \cite{AS}.     The    function
$M(-2\beta\theta,1-\beta,\beta^2z)$ diverges  exponentially for $z \to
\infty$, so must be rejected for $y>0$. Thus we write
\begin{eqnarray}
F_+(z)       &        =       &       A\,U\left(-2\beta\theta,1-\beta,
\frac{v_+\beta^2y^{1/\beta}}{Dx}\right)   \\     
F_-(z)    &    =    &
B\,U\left(-2\beta\theta,1-\beta,
\frac{-v_-\beta^2(-y)^{1/\beta}}{Dx}\right)    \nonumber     \\    &&+
C\,M\left(-2\beta\theta,1-\beta,\frac{-v_-\beta^2(-y)^{1/\beta}}{Dx}
\right),
\end{eqnarray}
with similar equations for $\tilde{F}_{\pm}(z)$ involving amplitudes 
$\tilde{A}$, $\tilde{B}$ and $\tilde{C}$.

Relations   between  the   coefficients   $A$,  $B$,   $C$,  and   the
corresponding  tilded variables   can be  obtained by  imposing the
boundary  conditions   (\ref{bc1})  and  (\ref{bc2}),   and  requiring
continuity of $Q$, $\partial  Q/\partial y$, $\tilde{Q}$ and $\partial
\tilde{Q}/\partial  y$  at  $y=0$.   After  some  straightforward  but
lengthy algebra,  these boundary and  continuity conditions eventually
lead to a consistency  condition on $\theta$, which determines $\theta
= \theta(p)$ as
\begin{equation}
\theta(p) = \frac{1}{4} - \frac{1}{2\pi\beta}\,
\sin^{-1}\left(\sqrt{\delta}\sin\left(\frac{\pi\beta}{2}\right)\right),
\label{general}
\end{equation}
where
\begin{equation}
\delta = \frac{2p^2\cos^2\left(\frac{\pi\beta}{2}\right) 
    + 2 \sinh^2\left(\frac{1}{2}\ln \gamma\right)}
    {\cos(\pi\beta) + \cosh(\ln\gamma)}
\end{equation}
and we recall that $\beta$, $\gamma$ are defined by Eq.\ (\ref{beta}).

The general result, Eq.\ (\ref{general}) can be checked in a number of 
special cases: \\

\noindent\underline{(i) $p=1$}. In this case $\delta=1$ and $\theta=0$. 
This is clearly correct since for $p=1$ there is no absorbtion and 
$Q(x,y,t)=1$. \\

\noindent\underline{(ii) $p=0$}. For this case, 
$\delta=(\sqrt{\gamma}-1/\sqrt{\gamma})^2/[\gamma+1/\gamma+2\cos(\pi\beta)]$. 
Inserting this into (\ref{general}), and carrying out some elementary 
manipulations, one recovers Eq.\ (\ref{result}), as required. \\

\noindent\underline{(iii) $\gamma=1$}. This gives $\delta=p^2$, and 
\begin{equation}
\theta = \frac{1}{4} - \frac{1}{2\pi\beta}\,
\sin^{-1}\left(p\,\sin\left(\frac{\pi\beta}{2}\right)\right).
\end{equation}
The  special case,  $\beta=1/3$,  of this  result  corresponds to  the
random acceleration problem  ($\alpha=1$) with partial absorption, and
gives $\theta = 1/4  - (3/2\pi)\sin^{-1}(p/2)$, recovering the results
of Burkhardt \cite{Burkhardt} and De Smedt et al.\cite{DeSmedt} for this 
special case.
In the following, we exploit the general result (\ref{general}) 
to obtain results for the crossing statistics  of the process defined 
by Eqs.\ (\ref{langevin1}) and (\ref{langevin2}). 

In  order to  place the  discussion  of crossing  statistics for  {\em
non-Gaussian}  processes in the  proper perspective,  we begin  with a
short discussion of crossing  statistics for {\em Gaussian} processes.
Within  the class of  models discussed  in this  paper, only  the case
$\alpha=1=\gamma$, corresponding  to the random  acceleration process,
is Gaussian.

For illustrative  purposes, we begin with the  random acceleration process,
$\dot{x}=y$,  $\dot{y}=\eta(t)$ or,  equivalently, $\ddot{x}=\eta(t)$,
with   $\langle   \eta(t)\eta(t')   \rangle  =   \delta(t-t')$.    For
convenience we  take the initial  condition $x(0)=0=\dot{x}(0)$.  Then
$x(t) =  \int_0^t dt'\ \int_0^{t'} dt''\ \eta(t'')$,  and the two-time
correlator is  $C(t_1,t_2) = <x(t_1)(x(t_2)> =  t_2t_1^2/2 - t_1^3/6$,
where  we have taken  $t_2 \ge  t_1$ without  loss of  generality. The
normalized two-time correlator is
\begin{equation}
\tilde{C}(t_1,t_2)= \frac{C(t_1,t_2)}{\sqrt{C(t_1,t_1)C(t_2,t_2)}} = 
\frac{3}{2}\left(\frac{t_1}{t_2}\right)^{1/2} 
- \frac{1}{2}\left(\frac{t_1}{t_2}\right)^{3/2},
\label{tscaling}
\end{equation}
for  $t_2  \ge  t_1$.  Notice  that  $\tilde{C}(t_1,t_2)$  depends  on
$t_1,t_2$ only through the {\em ratio} $t_1/t_2$. This is an immediate
consequence   of   the  fact   that,   with   the  initial   condition
$x(0)=0=\dot{x}(0)$,  there  is  no   timescale  in  the  problem,  so
dimensionless  correlation  functions can  only  depend  on ratios  of
times. If the initial position  and velocity are non-zero, the scaling
form (\ref{tscaling}) is  recovered in the limit that  $t_1$ and $t_2$
are both  taken large with  the ratio held  fixed. It should  be noted
that this scaling property  of temporal correlations is not restricted
to Gaussian processes, but holds for all the models discussed here.

This scaling  property implies that,  if one introduces  a logarithmic
time variable \cite{MS,MSBC}, $T=\ln t$, normalized correlation functions
can only depend on {differences} of $T$-variables, i.e.\ the processes
become {\em  stationary} in logarithmic time.  Again,  this applies to
both  the Gaussian and  the non-Gaussian  processes that  we consider.
For  the  random  acceleration  process, it  follows  immediately  from
(\ref{tscaling}) that  the normalized correlator  in logarithmic time,
$f(T)  = \langle X(T)\,X(0)\rangle$  where $X(T)  = x(t)/\sqrt{\langle
x^2(t)}$, is
\begin{equation}
f(T) = (3/2)\exp(-T/2)-(1/2)\exp(-3T/2)
\label{randaccn}
\end{equation}
for  $T \ge 0$  (and $f(-T)=f(T)$).


For Gaussian  processes there is a major  simplification: the two-time
correlator  implicitly  determines   all  properties  of  the  system,
including  the  first-passage   exponent  $\theta$  and  the  crossing
statistics. As an example, consider  the mean density, $\rho$, of zero
crossings in  logarithmic time. The mean density $\rho$ is finite only 
for {\it smooth} Gaussian processes whose correlator $f(T)$ has the 
short time behavior, $f(T)=1- a T^2$ as $T\to 0$
where $a=-f''(0)/2$ is finite. For such processes, the  mean number of  
zero-crossings of the process $X(T)$ in time interval $T$ is
\begin{equation}
\langle n \rangle = \int_0^T dT' \langle \delta(X(T'))|\dot{X}(T')|\rangle
= T \langle \delta(X(T')) \rangle\,\langle |\dot{X}(T')|\rangle
\end{equation}
since $X(T)$  and $\dot{X}(T)$  are uncorrelated for smooth processes,
$\langle X(T) \dot{X} (T)\rangle = f'(0)=0$. Furthermore,  $X$ and
$\dot{X}$  are  normally   distributed  with  variances  $f(0)=1$  and
$-f''(0)$ respectively.  It follows that the mean crossing density is
given by 
\begin{equation}
\rho = \langle n \rangle/T = \frac{1}{\pi}\sqrt{-f''(0)},
\label{firstmomentG}
\end{equation}
a result first derived by Rice \cite{Rice}.  

In a similar way it is possible to obtain an (albeit rather more complicated) 
expression for the variance, $\langle n^2 \rangle -\langle n \rangle^2$, of 
the number of crossings in time $T$, again expressed in terms of  the 
correlator $f(T)$, a result due to Bendat~\cite{Bendat}.

For non-Gaussian processes, this approach no longer works. The first line, 
$\langle n \rangle = T \langle \delta(X)\,|\dot{X})|\rangle$ still holds, 
but since the stationary distribution of $(X,\dot{X})$ is in general not 
known, further progress seems to be impossible. We will show, however, that 
exact results can be obtained for the class of non-Gaussian models discussed 
in this paper by exploiting our knowledge of the function $\theta(p)$. The 
connection between the two was first noted in ref.\cite{Majumdar-Bray}. 

Working in logarithmic time, as discussed above, we can write the survival 
probability $Q(T)$ for the partial survival problem in the form
\begin{equation}
Q(T) = \sum_{n=0}^\infty p^n P_n(T)
\label{Q(T)}
\end{equation}
where $P_n(T)$ is the probability of the process crossing the line $x=0$  
$n$ times in the (logarithmic) time interval $T$, and $p^n$ is the probability 
of surviving all $n$ crossings. Now $Q(T)$ decays asymptotically as 
$Q(T) \sim \exp(-\theta(p)T)$, and the right-hand side of Eq.\ (\ref{Q(T)}) 
can be written in terms of the cumulants of $n$, to give
\begin{equation}
\exp \left(\sum_{r=0}^\infty \frac{(\ln p)^r}{r!}\langle n^r \rangle_c\right)
\sim \exp\left(-\theta(p)\,T\right),
\end{equation}
for large $T$, where $\langle n^r \rangle_c$ is the $r^{\rm th}$ cumulant 
of $n$, and therefore
\begin{equation}
\sum_{r=0}^\infty \frac{(\ln p)^r}{r!}\langle n^r \rangle_c 
\sim -\theta(p)\,T,
\label{expansion}
\end{equation}
for large $T$.

The cumulants of $n$ can now  be determined by expanding both sides of
Eq.\  (\ref{expansion}) around  $p=1$, by  writing  $p=1-\epsilon$ and
equating  coefficients   of  powers  of  $\epsilon$,   to  obtain  the
cumulants of $n$ in terms of the derivatives of $\theta(p)$ evaluated
at $p=1$. In this way one obtains
\begin{eqnarray}
\langle n \rangle & = & -\theta'(1) T \\
\langle n^2 \rangle_c - \langle n \rangle & = & - \theta''(1) T
\end{eqnarray}
etc. This approach gives, for the mean crossing density,
\begin{equation}
\rho = \frac{\langle n \rangle}{T} = \frac{1}{2\pi\beta}\,
\frac{\sin(\pi\beta)}{\cos(\pi\beta) + \cosh(\ln \gamma)}.
\label{firstmoment}
\end{equation}

For the special case $\beta=1/3$, $\gamma=1$ that corresponds to the 
(Gaussian) random acceleration process, we can check Eq.\ (\ref{firstmoment}) 
against the general result (\ref{firstmomentG}) that holds for any Gaussian 
stationary process. We find that both expressions reduce to 
$\rho=\sqrt{3}/2\pi$ for this case. 

In a similar way we can calculate the second cumulant of the number 
of crossings by expanding to second order in $1-p$. The result for 
$\langle n^2 \rangle_c \equiv \langle n^2 \rangle - \langle n \rangle^2$ is 
\begin{equation}
\frac{\langle n^2 \rangle_c}{T} = \frac{1}{2\pi\beta}\,[2F \sin(\pi\beta) 
- F^2 \sin(2\pi\beta)],
\label{secondcumulant}
\end{equation}
where
\begin{equation}
F = 1/[\cos(\pi\beta) + \cosh(\ln \gamma)].
\end{equation}
Again, this result can be checked for the random acceleration process 
$\beta = 1/3$, $\gamma = 1$, which is Gaussian. For this case, Eq.\ 
(\ref{secondcumulant}) gives $\langle n^2 \rangle_c/T =2/\pi\sqrt{3}$.
On the other hand, the second cumulant can be calculated exactly for 
any Gaussian process using Bendat's formula. This gives the result 
$2/\pi\sqrt{3}$, as expected. 

It is noteworthy that for the class of models discussed here, arbitrary
cumulants  of  the crossing  number  are  readily  obtained by  simply
computing  the appropriate  derivatives of  $\theta(p)$,  evaluated at
$p=1$. Thus a knowledge of $\theta(p)$ provides a powerful tool.  Even
for Gaussian  processes, a general  expression for the moments  of $n$
above the second is not  available for models where $\theta(p)$ is not
known,   which  is   essentially  all   but  a   few   special  models
\cite{Majumdar-Bray}.


It is also interesting to compute the mean time intervals (logarithmic
scale)  $l_{\pm}$ between crossings  that the  particle spends  on the
positive  (negative) side  of  the  $X$ axis.   For  the special  case
$\gamma=1$ (where $v_+=v_{-}$),  it is clear that $l_{+}=l_{-}=1/\rho$
with  $\rho$  being the mean density  of  crossings  given  by  Eq.\ 
(\ref{firstmoment}).   However, for  arbitrary $\gamma\ne  1$, $l_{+}$
is, in  general, different from  $l_{-}$. Let $\rho_{\pm}$  denote the
mean number  of crossings of  $X=0$ from the  right (left) of  the $Y$
axis.   Clearly  $\rho_+=\rho_{-}=\rho/2$.   To  calculate   the  mean
intervals  $l_{\pm}$  in the  general  case  ($\gamma\ne  1$) one  can
proceed as  follows.  We consider the normalized  process $X(\tau)$ in
the logarithmic  time scale $\tau$ so  that it is  stationary.  Let us
first define a new variable,  the `occupation time', that measures the
fraction of time spent by  the process $X(\tau)$ above (below) the $X$
axis,  $L_{\pm}=  \frac{1}{T}\int_0^T  \theta(\pm  X(\tau))\,  d\tau$.
Taking  the average, and using the stationarity,   one  gets  $\langle
L_{\pm}\rangle = \langle \theta(\pm X)\rangle$. However, it is evident
that $\langle  L_{\pm}\rangle= \rho_{\pm}  l_{\pm}$.  Hence we  get an
expression for the mean intervals
\begin{equation}
l_{\pm} = \frac{1}{\rho_{\pm}} \langle \theta(\pm X)\rangle.
\label{meanintervals}
\end{equation}
For  the  case where  one  has  the symmetry  $X\to  -X$, such as  the
case $\gamma=1$, one gets, using $\langle \theta(X)\rangle  =1/2$, the
expected  result $l_+=l_{-}=1/\rho$. However,  for $\gamma\ne  1$, the
exact knowledge  of $\rho$ from Eq. (\ref{firstmoment})  is not enough
to  calculate the  mean  size  of intervals  $l_{\pm}$.  One needs  to
compute, in addition, the  quantity $\langle \theta(\pm X)\rangle$. To
calculate  this average  we need  to know  the  stationary probability
density  $P(X)$,  since  $\langle \theta(X)\rangle  =  \int_0^{\infty}
P(X)\, dX$.

The calculation of the  probability density $P(X)$, for $\gamma\ne 1$,
is nontrivial.  The only case  where we have succeeded  in calculating
$P(X)$  exactly  is the  case  where  $\alpha=0$.  In this  case,  the
equation  of  motion $x(t)$  in  the  original  time $t$   reads, from
Eq.  (\ref{langevin2}), $\dot{x}=v_+\theta(y)-v_{-}  \theta(-y)$. Then
one can  write, $x(t) =  \epsilon t +  v T_t $ where  $T_t= \int_0^{t}
{\rm sign}[y(t')]\,dt'$ is the sign-time of an ordinary Brownian motion,
$\epsilon=(v_{+}-v_{-})/2$     measures     the     anisotropy     and
$v=(v_{+}+v_{-})/2$.  The distribution of $T_t$ for Brownian motion is
well  known to  have the  famous arcsine  form  of L\'evy~\cite{Levy},
$P(T_t,  t) =  \frac{1}{t}f\left(\frac{T_t}{t}\right)$  where $f(x)  =
1/[\pi \sqrt{1-x^2}]$  for $-1\le x\le 1$ and  $f(x)=0$ outside. Using
this   result,   one   gets   the  exact   distribution   of   $x(t)$,
$P(x,t)=\frac{1}{vt}G\left(\frac{x}{vt}\right)$  where  $G(z)=  1/[\pi
\sqrt{1-(z-\epsilon)^2}]$ for $-1+\epsilon/v\le z\le 1+\epsilon/v$ and
$G(z)=0$ otherwise. Carrying out the  integral of $P(x,t)$ only over the
positive   (negative)   $x$  axis,   one   gets  $\langle   \theta(\pm
x(t))\rangle     =     \frac{1}{2}\pm    \frac{1}{\pi}     {\sin}^{-1}
\left(\frac{\epsilon}{v}\right)$.   Using   $\theta(X)=\theta(x)$  and
$\rho= 1/[\pi \cosh(\ln \gamma)]$  (obtained by putting $\beta=1/2$ in
Eq. (\ref{firstmoment}))  we finally get  the exact mean intervals for
the $\alpha=0$ case:
\begin{equation}
l_{\pm}  = 2\pi  \cosh(\ln \gamma)  \left[\frac{1}{2}\pm \frac{1}{\pi}
{\sin}^{-1}\left(\frac{v_{+}-v_{-}}{v_{+}+v_{-}}\right)\right].
\label{meanintervals2}
\end{equation}
The determination of $l_{\pm}$ for other values of $\alpha$ (including
the $\alpha=1$ case) remains an open problem.

In  this  paper we  have  calculated  the  partial survival  exponent,
$\theta(p)$,  for  a class  of  (in  general non-Gaussian)  stochastic
processes describing a random  walker moving in a transverse ``shear''
flow. We  have then  used the result  for $\theta(p)$ to  derive exact
expressions  for the  first  two cumulants  of  the crossing  number
(number  of crossings  of the  line  $x=0$) working  on a  logarithmic
timescale where the process is stationary.  To our knowledge these are
the  first results  for the  statistics  of crossing  numbers for  any
non-Gaussian process. We have checked that our general result reduces,
in special  cases, to  the known results  for the  random acceleration
process. We have also computed exactly the mean time intervals between
successive   crossings  of  the   $y$  axis   for  the   special  case
$\alpha=0$. The calculation of  the mean time intervals for $\alpha\ne
0$ remains a challenging open problem.


\end{document}